\documentclass[aps,showkeys,superscriptaddress,byrevtex,a4paper,12pt]{revtex4}
\usepackage{dcolumn}
\usepackage{bm}
\usepackage[latin2]{inputenc}
\usepackage{graphics}
\usepackage{longtable}
\usepackage{amsmath}
\usepackage{amssymb}
\usepackage{enumerate}

\begin{document}


\title{Methods for electronic-structure calculations - an overview from a reduced-density-matrix point of view  }

\author{P. Ziesche}
\email[E-mail: ]{pz@mpipks-dresden.mpg.de}
\affiliation{Max-Planck-Institut f\"ur Physik komplexer Systeme,
  N\"othnitzer Str. 38, D-01187 Dresden, Germany}
\date{\today}

\author{F. Tasn\'adi}
\email[E-mail: ]{f.tasnadi@ifw-dresden.de}
\affiliation{Leibniz-Institut f\"ur Festk\"orper- und Werkstoffforschung, \\
Helmholtzstr. 20, D-01069 Dresden, Germany, and \\
University of Debrecen, Hungary}


\begin{abstract}
\noindent
The methods of quantum chemistry and solid state theory to solve the many-body
problem are reviewed. We start with the
definitions of reduced density matrices, their properties (contraction sum
rules, spectral resolutions, cumulant expansion, $N$-representability), and
their determining equations (contracted Schr\"odinger equations) and we
summarize recent extensions and generalizations of the traditional quantum
chemical methods, of the density functional theory, and of the quasi-particle
theory: from finite to extended systems
(incremental method), from density to density matrix (density matrix functional
theory), from weak to strong correlation (dynamical mean field theory), from
homogeneous (Kimball-Overhauser approach) to inhomogeneous and finite systems.
Measures of the correlation strength are discussed.
The cumulant two-body reduced density matrix proves to be a key quantity. Its
spectral resolution contains geminals, being possibly the solutions of an
approximate effective two-body equation, and the idea is sketched of how its
contraction sum rule can be used for a variational treatment.

\end{abstract}


\maketitle

\section{Introduction}
\label{section1}

In the electron theory of atoms, molecules, clusters, and solids ($N$ electrons are confined
to a volume $\Omega$) the pairwise
Coulomb repulsion between electrons together with the Pauli `repulsion' between
electrons with parallel spins causes the complex phenomenon of electron
correlation \cite{Ful1}-\cite{con4}. The problem is, to calculate this electron
correlation from first principles. Even within the simplifying non-relativistic
description and within the Born-Oppenheimer approximation, this problem can not
be solved exactly. In some cases, correlation beyond the Hartree-Fock approximation
causes only quantitative changes, but there are other cases, where it makes qualitative
changes. So, electron correlation let exist the dimers F$_2$ and Hg$_2$, or it gives the 
correct sign for the dipole moment of CO. Electron correlation allows a solid
to be a metal (in the Hartree-Fock approximation the density of states vanishes at the
Fermi energy according to $\sim 1/\ln|\varepsilon-\varepsilon_{\rm F}|$). Correlation brings 
the Curie temperature of ferromagnets down to the experimental values (in the Hartree-Fock
approximation they are much too high). 
Electron correlation is important for the ground states of transition-metal oxides to be 
insulating or metallic, ferro- or antiferro-magnetic. Correlation has its peculiarities 
or needs different computational methods for closed-shell systems and open-shell systems, 
for finite and extended systems, for systems near and far from the equilibrium (defined 
by vanishing Hellmann-Feynman forces), for the ground state (GS)
and for excited states (for a solid, this means its quasi-particle band structure), 
for different types of solids,
for weak or moderate correlation (as for sp metals and semiconductors) and
strong correlation (as for systems with open d-shells or open f-shells, e.g.,
transition metal oxides and rare earth or actinide systems, cf. Appendix). Here we focus
on the GS. Its correlation shows up in the many-body (MB) wavefunction (WF) of the GS (e.g.
in a Jastrow factor in front of a Slater determinant) or in the correlation energy (being
unfortunately not sensitive to distinguish between weak and strong electron correlation)
or in quantum-kinematical quantities, which follow from the MB WF by integration over most of
the variables leaving only a few of them unintegrated. Thus hierarchies of reduced densities
(RDs) $\rho_1,\rho_2,\cdots$ and reduced density matrices (RDMs) $\gamma_1,\gamma_2,\cdots$
arise with the advantage to allow a distinction between
weak and strong correlation. Therefore they are used to define
measures of the correlation strength. One possibility is the correlation induced
non-idempotency of the 1-body RDM (1-matrix) $\gamma_1$. Another possibility
is to derive correlation influenced particle-number fluctuations from the pair density (PD)
$\rho_2$ in parts of the system; the conclusion is: correlation suppresses such
fluctuations \cite{Ful1}, p. 157, \cite{Zie8}.

Important issues of the RDM theory are listed in the following: the
contraction properties of
the RDMs $\gamma_1, \gamma_2, \cdots$ and their cumulant expansions (Sec.2),
parts of the total energy as
functionals of RDs or RDMs (Sec.3), the hierarchy of contracted Schr\"odinger equations (CSEs)
[what means the equation for $\gamma_1$ contains $\gamma_2$ and $\gamma_3$, the equation for
$\gamma_2$ contains $\gamma_3$ and $\gamma_4$, $\cdots$] (Sec.4), and the
size-extensivity [because only size-extensive quantities are compatible with
the thermodynamic limit (TDL), which means $N,\Omega\to\infty, N/\Omega={\rm const}$] (Sec.5).
Finally a comprehensive summary of recent developments in the calculational
treatment of the many-electron problem is given (Sec.6). Thereby, ``old''
methods to calculate the electronic structure like the WF based
methods [namely, configuration interaction (CI), coupled cluster (CC),
M\o ller-Plesset (MP), and quantum Monte Carlo (QMC) for finite systems] and
the density based density
functional theory (DFT) \cite{Hoh,Levy} and the Green's function based
quasi-particle theory (QPT) \cite{Hed} for both finite and extended systems are
compared with some ``new'' methods like the WF based incremental method
\cite{Grae,Alb,Ful3} (thereby the nearsightedness of electron correlation
\cite{Kohn} is used) or the 1-matrix based density-matrix functional theory
(DMFT) (being conceptionally simpler and yielding more informations than the
DFT) \cite{Gil,Cio2,Cio3,Herb1}, and also extensions towards strong correlation
(Sec.VI(c),(i), Appendix). Within the RDM theory \cite{Dav1,Erd,Cio1,Cole},
there seem to be three possibly prospective (?) developments: (i) the
$N$-representability studies (Kummer variety, Coleman's algorithm, linear
inequalities) \cite{Cole}, (ii) the CSE method for finite
systems \cite{Cio1}, p. 85, 117, 139, \cite{Nakuji,Nak,Mazz} including critical
discussions of the thereby used `reconstructions' \cite{Harr,Herb,Noo} and alternative
suggestions (Kutzelnigg \cite{Noo}), and
(iii) the Kimball-Overhauser approach for the PD of the homogeneous electron
gas (HEG) in terms of 2-body WFs (geminals), cf. \cite{Kim,Davou,Zie5} and
refs. therein. - The paper contains also open questions with the intention to
stimulate further research.

\begin{table}[h!]
\caption{\label{tab:1st} List of acronyms}
\begin{tabular}{|l|l|l|l|}
\hline
AGP & antisymmetrized geminal power & MB & many-body \\
CC & coupled cluster &  MP & M\o ller-Plesset\\
CI & configurational interaction  & MR & multi reference \\
CSE & contracted Schr\"odinger equation & PDFT & pair density functional
theory \\
DFT & density functional theory & QMC  & quantum Monte Carlo\\
DMFT & density matrix functional teory & QPT & quasi-particle theory \\
ELF & electron localization function & RD & reduced density \\
GGA & generalized gradient approximation & RDM & reduced density matrix \\
GS & ground state & SR & sum rule \\
HEG & homogeneous electron gas& TDL & thermodynamic limit \\
KS & Kohn-Sham & XC & exchange-correlation \\
LDA & local density approximation & WF & wave function \\
\hline
\end{tabular}
\end{table}

\section{Basic quantum kinematics}
\label{section2}

The GS of a non-relativistic $N$-electron system in
Born-Oppenheimer approximation is described by a $N$-body WF $\Psi(1,\cdots,N)$
with $1=(\mathbf{r}_1,\sigma_1)$ and $\int d1=\int d^3r_1\sum_{\sigma_1}$ or
equivalently by a hierarchy of hermitian RDMs $\gamma_1, \gamma_2, \gamma_3,
\cdots$ [first introduced by K. Husimi \cite{Hus}] with the following
contraction sum rules (SRs)
\begin{equation}
\gamma_1(1|1')=\int \frac{d2}{N-1}\ \gamma_2(1|1',2|2), \quad
\gamma_2(1|1',2|2')=\int \frac{d3}{N-2}\ \gamma_3(1|1',2|2',3|3), \cdots
\label{gammas}
\end{equation}
or $\text{C}\gamma_2=(N-1)\gamma_1, \text{C}\gamma_3=(N-2)\gamma_2, \cdots$ for
short, i.e. the contraction operator C makes the last variable pair diagonal ($i'=i$) and
integrates it ($\int di$).
This chain $\gamma_N(=\Psi\Psi'^*)\rightarrow\cdots\rightarrow\gamma_3
\rightarrow\gamma_2\rightarrow\gamma_1$ is called $N$-representability, 
$\Psi'^*=\Psi^*(1',\cdots,N')$. It
becomes infinitely long in the TDL. Note ${\rm Tr}\gamma_1=N$, ${\rm Tr
\gamma_2}=N(N-1), \cdots$, ${\rm Tr}\gamma_N=N!$. The diagonal elements of
$\gamma_1, \gamma_2, \gamma_3,\cdots$ are the density $\rho_1$, the pair
density (PD) $\rho_2$ [containing the Fermi hole for the spin-parallel electron
pairs and the Coulomb hole for the spin-antiparallel electron pairs], the
triple density $\rho_3$, $\cdots$:
\begin{equation}
\rho_1(1)=\gamma_1(1|1), \quad \rho_2(1,2)=\gamma_2(1|1,2|2), \quad
\rho_3(1,2,3)=\gamma_3(1|1,2|2,3|3), \cdots
\end{equation}
or $\rho_1=\text{D}\gamma_1, \rho_2=\text{D}\gamma_2, \rho_3=\text{D}\gamma_3,
\cdots$, i.e. the diagonal operator D makes all the variable pairs diagonal (1'=1, 2'=2, $\cdots$).
Note $\rho_i\ge 0$. If $X$ is a certain part of the system (e.g. a
Daudel loge, refs. on this loge theory are in Ref.\cite{Zie12,Nal}), then
\begin{equation}
N_{X}=\int_{X}d1\ \rho_1(1), \;
\binom{N}{2}_{X}=\int_{X}\frac{d1d2}{2!}\ \rho_2(1,2),
\cdots
\end{equation}
have respectively the meaning of the average number of particles in $X$,
the average number of particle pairs in $X$ etc., approaching $N$, $\binom{N}{2}$, 
$\cdots$ for $X\to\infty$. The term `average number of particles in 
$X$' means: the number of particles in
$X$ fluctuates. In lowest order, such fluctuations are described by the mean
squared deviation $\Delta N_{X}$. They are determined by the 2-body function
$w_2(1,2)$ [defined by $\rho_2(1,2)=\rho_1(1)\rho_1(2)-w_2(1,2)$] according to
\cite{Zie8}
\begin{equation}
\frac{(\Delta N_{X})^2}{N_{X}}=1-\frac{1}{N_{X}}\int_{X}d1\int_{X}d2
\ w_2(1,2), \quad \int d1\int d2 \ w_2(1,2)=N.
\label{2bodyw}
\end{equation}
$w_2(1,2)$ is size-extensively normalized. Particle-number fluctuations can be 
described in more detail by $P_{X}(M)$, the probability of finding $M(=0,1,\cdots,
N)$ particles in $X$ with
\begin{equation}
\sum_{M=0}^{N}P_{X}(M)=1, \quad \sum_{M=0}^{N} P_{X}(M)M=N_{X}, \quad
\sum_{M=0}^{N}P_{X}(M)\left( M-N_{X}\right)^2=(\Delta N_{X})^2.
\end{equation}
The example \cite{Zie8}
\begin{equation}
P_{X}(1)= \int_{X}d1 \ \rho_1(1) -
\frac{1}{1!} \int_{X}d1\int_{X}d2 \ \rho_2(1,2) +
\frac{1}{2!}\int_{X}d1\int_{X}d2\int_{X}d3 \ \rho_3(1,2,3)- \cdots
\label{PX1}
\end{equation}
shows, how e.g. $P_X(1)$ follows from the reduced densities. For other values of 
$M$ cf. \cite{Zie8}. An approximate expression needs only $\rho_1$ and $\rho_2$, 
namely $P_{X}(M)\sim {\rm e}^{-\beta(M-\alpha)^2}$
with distribution parameters $\alpha$ and $\beta$ determined by $N_{X}$ and
$\Delta N_{X}$. [This expression maximizes the entropy
$-\sum_M P_{X}(M)\ln P_{X}(M)$ under the
constraints of given $N_{X}$ and $\Delta N_{X}$.] The fluctuation analysis of
finite and extended systems based on this approximation shows that `(strong)
correlation (strongly) suppresses fluctuations' \cite{Ful1,Zie8}; for
attractive
interaction (of the Calogero-Sutherland model = one-dimensional Fermi gas with
$1/x_{ij}^2$ interaction), they are enhanced \cite{Roe}. For the peculiarities
of strong electron correlations cf. Appendix. - The `on-top' curvature of the
Fermi hole is used as a local measure of the correlation strength \cite{Zie12}. 
It is also derived an `electron localization function (ELF)'
\cite{Becke}; an alternative is introduced in Ref.\cite{Kohout}.

Because the RDMs are hermitian, they can be diagonalized with
\begin{equation}
\gamma_1(1|1')=\sum_{\kappa} \psi^{}_{\kappa}(1)\nu_{\kappa}\psi^{\ast}_{\kappa}(1'), \quad
0<\nu_{\kappa}<1, \quad \sum_{\kappa}\nu_{\kappa}=N
\label{spectral1}
\end{equation}
as the simplest example. The $\psi^{}_{\kappa}(1)$ are called natural orbitals
(more precisely spin-orbitals) with natural occupancies $\nu_{\kappa}$. For the
HEG, the $\psi_\kappa(1)$ are plane waves and the $\nu_\kappa$ yield the
momentum distribution $n(k)$, recently parametrized in Ref.\cite{Zie11}. The
occupancies are only exceptionally idempotent, in general they are
non-idempotent: $\nu_{\kappa}^2<\nu_{\kappa}$ and $\sum_{\kappa}
\nu_{\kappa}^2<N$. Because P.-O. L\"owdin has asked for the meaning of
$\sum_{\kappa}\nu_{\kappa}^2=\text{Tr}\gamma_1^2$ \cite{Low}, the quantity
\begin{equation}
c_2=1-\text{tr}\gamma_1^2=1-\frac{1}{N}\sum_{\kappa}\nu_{\kappa}^2, \quad
0<c_2<1
\label{Low}
\end{equation}
will be referred to as L\"owdin parameter (with
$\text{tr}=\frac{1}{N}\text{Tr}$). For non-interacting single-reference cases
(only one Slater determinant) and therefore
idempotent occupancies $\nu_{\kappa}=0$ or $1$, this parameter vanishes.
Interaction then causes a correlation tail
in the CI expansion with a lot of additional Slater determinants beyond the
leading Slater determinant. This makes the $\nu_\kappa$ non-idempotent, thus
$c_2>0$. So, $c_2$ is another measure of the correlation strength
\cite{Zie9,Zie10}. An increasing weight of the CI correlation tail increases
also $c_2$ (with perhaps weak, moderate, and strong correlation for $c_2\ll
1/2$, $c_2\approx 1/2$, and $c_2\gg 1/2$, respectively). Thereby the CI expansion is
thought to be unique (basis set independent) by representing it through natural
spin orbitals.  Often one has to start
with a multi-reference (MR) state, a sum of several Slater determinants
to correctly describe, e.g., a dissociation
limit or nearly degeneracy. Then the L\"owdin parameter is
already non-zero for `no CI correlation tail' ($c_{\text{MR}}>0$. For this
description the term `non-dynamical or static correlation' is used) following
Clementi and Corongiu \cite{Clem}. Full CI with its additional correlation tail
(`dynamical correlation') makes then $c_2>c_{\text{MR}}$. -
For a solid, $\nu_\kappa$ contains the occupancy band structure \cite{Koch}. -
The 2-body analog of Eq.(\ref{spectral1}) is the spectral resolution of the
2-matrix $\gamma_2$, containing antisymmetric 2-body WFs (natural geminals).
Are they the solutions of a 2-body Schr\"odinger equation with an appropriately
screened Coulomb repulsion (similar as this is the case with the Kimball-Overhauser
approach for the HEG \cite{Kim})?

An important aspect of the quantum kinematics is the cumulant expansion of RDMs
and RDs \cite{Kla}. Its most compact representation uses generating
functionals, cf. e.g. P. Ziesche in \cite{Cio1}, p. 33. For the simplest case
of the 2-matrix this means
\begin{equation}
\gamma_2(1|1',2|2')=\gamma_1(1|1')\gamma_1(2|2')-\gamma_1(1|2')\gamma_1(2|1')-\chi_2(1|1',2|2')
\label{CumulantExp}
\end{equation}
or $\gamma_2=\text{A}\gamma_1\gamma_1-\chi_2$ for short (A = antisymmetrizer).
The factorized (or
disconnected) terms are the generalized (because the natural occupancies are
non-idempotent) Hartree-Fock part of $\gamma_2$ with $\text{Tr}
\gamma_2^{\text{HF}}=N(N-1)+Nc_2$. The non-factorizable remainder $\chi_2$ is
called cumulant 2-matrix. In perturbation theory, $\chi_2$ is given by
(size-extensive) linked diagrams; therefore it is also called connected part of
$\gamma_2$. [An equation similar to Eq.(\ref{CumulantExp}) defines the cumulant
part of the 2-body Green's function.] The above mentioned contraction SR
$\text{C}\gamma_2=(N-1)\gamma_1$
(which is not size-extensive) is transformed to $\text{C}\chi_2=\gamma_1-
\text{C}\gamma_1^2$ or
\begin{eqnarray}
\int d2 \ \chi_2(1|1',2|2)&=&\gamma_1(1|1')-\int d2 \ \gamma_1(1|2)
\gamma_1(2|1')
=\sum_{\kappa}\psi^{}_{\kappa}(1) \nu_{\kappa}(1- \nu_{\kappa})
\psi^{\ast}_{\kappa}(1').
\label{Contraction1}
\end{eqnarray}
The invariance of the rhs under the exchange $\nu_{\kappa}\leftrightarrow
(1- \nu_{\kappa})$ is called particle-hole symmetry \cite{Rus}. 
Eq.(\ref{Contraction1}) contains the normalization SR
\begin{eqnarray}
\text{tr}\chi_2=\frac{1}{N}\sum_{\kappa}\nu_{\kappa}(1- \nu_{\kappa})=c_2;
\quad s= -\frac{1}{N}\sum_\kappa[\nu_\kappa\ln \nu_\kappa+
(1-\nu_\kappa)\ln(1-\nu_\kappa)].
\end{eqnarray}
Thus the  L\"owdin parameter
$c_2$ measures the non-idempotency of the natural occupancies $\nu_{\kappa}$
and simultaneously the normalization of the cumulant 2-matrix $\chi_2$ per
particle, thereby not affecting the normalization of the total 2-matrix
$\gamma_2$. $s$ defines the particle-hole symmetric correlation entropy \cite{Zie11,Zie9}.
For the PD $\rho_2$ the cumulant expansion Eq.(\ref{CumulantExp}) means
\begin{equation}
\rho_2(1,2)=\rho_1(1)\rho_1(2)-\gamma_1(1|2)\gamma_1(2|1)-u_2(1,2), \quad
u_2(1,2)=\chi_2(1|1,2|2).
\label{rho2}
\end{equation}
$u_2$ may be addressed as cumulant PD. Comparison with Eq.(\ref{2bodyw})
shows $w_2(1,2)=\gamma_1(1|2)\gamma_1(2|1)+u_2(1,2)$, being thus the exchange
including cumulant PD.

If $\chi_2$ is available from perturbation theory or from some effective 2-body
theory, then the SR Eq.(\ref{Contraction1}) allows one to
determine the 1-matrix $\gamma_1$ by means of diagonalizing
Eq.(\ref{Contraction1}). The resulting quadratic equation
$\lambda_\kappa=\nu_\kappa (1-\nu_\kappa)$ is solved by $\nu_\kappa^\pm=
\frac{1}{2}\pm{\sqrt {\frac{1}{4}-\lambda_\kappa}}$.
Associating certain $\kappa$'s to $+$ and all the others to $-$,
yields occupancies $\nu_{\kappa}>\frac{1}{2}$ and $\nu_{\kappa}<\frac{1}{2}$
respectively, such that $\sum_\kappa \nu_\kappa=N$. [For the low-density HEG($r_s\gg 1$),
i.e. for the Wigner crystal only $\nu_\kappa^-$ has a physical meaning and it is
$\lambda_\kappa\ll 1$, thus $\nu_\kappa^-\approx\lambda_\kappa$.]
Let us denote the described
procedure by $\gamma_1=\tilde{\text{C}}\chi_2$. With this $\gamma_1$ the exact
MB kinetic energy $T$ and the L\"owdin parameter $c_2$ can be
calculated and also the PD $\rho_2$, which is needed for the interaction energy
$V_{\rm int}$ and for the fluctuation analysis (both in direct and momentum
space). So, the cumulant 2-matrix $\chi_2$ is the most interesting quantum
kinematic quantity. Although it is a key quantity, unfortunately not much is known
about it. Questions:
\begin{itemize}
\item What is known about the spectral resolution of the cumulant 2-matrix $\chi_2$?
\begin{eqnarray}
\chi_2(1|1',2|2')=\frac{1}{2!}\sum_K \psi^{}_K(1,2) \mu_K \psi^{\ast}_K(1',2'),
\quad \frac{1}{N}\sum_K \mu_K =c_2.
\end{eqnarray}
Which values are possible for the cumulant occupancies $\mu_K$?
\item $\gamma_2$ is
$N$-representable \cite{Cole}. What are the consequences for $\chi_2$?
\item What are the peculiarities of $\chi_2$ for metals (for the
high-density HEG cf. \cite{Zie1}), quasi-one-dimensional charge-density-wave 
conductors, semiconductors, (band, Peierls, 
charge-transfer, Mott-Hubbard) insulators, superconductors
(off-diagonal-long-range order), ferro/antiferro-magnets, ferroelectrica?
For cases of strong electron correlation cf. Appendix.
\item What are the peculiarities of $\chi_2$ for cases, where one of the
occupancies of $\gamma_2$ is very large \cite{Cole}. There exists a rich
literature on antisymmetrized geminal power (AGP) WFs, cf. \cite{Erd,Ort}.
M. Rosina has studied the 2-matrix $\gamma_2$ of the AGP WF \cite{Erd2}. The
AGP function is a flexible ansatz for fermion systems with arbitrary $N$
\cite{Cole}. Extreme AGP WFs have (for large $N$) very large 2-matrix occupancies
equaling $[N/2]$ \cite{Erd}, p. 35, \cite{Yang}. The GS of Be has a large 2-matrix 
occupancy ($>1$) (V.H. Smith, priv. commun.). What does all this mean for the
cumulant occupancies $\mu_K$?
\item Electron correlation is discussed in terms of
its nearsightedness \cite{Kohn}. How does this show up by $\chi_2$?
\end{itemize}
A drawback of $\chi_2$ is that the positivity condition $\rho_2\geq 0$ becomes
more complicated.

The normalization and contraction of $\chi_3$ yield
${\rm Tr}\chi_3=2\sum_\kappa \nu_\kappa(1-\nu_\kappa)(1-2\nu_\kappa)\equiv Nc_3$
and ${\rm C}\chi_3(1|1',2|2',3|3')=2\chi_2(1|1',2|2')-
[{\rm A'}{\rm C}\chi_2(1|1',2|3)\chi_1(3'|2')+ {\rm h.c.}]$, where the
contraction operator C makes $3'=3$ and $\int d3$ as defined above, $\rm A'$
is the antisymetrizer with respect to the primed variables and h.c. means
hermitian conjugate. Thereby integrals $\int d3 \psi_K(1,3)\psi_\kappa^*(3)$
appear, which describe the overlap of a natural orbital with a cumulant geminal.


\section{Basic functionals}
\label{section3}

Explicitly defined functionals for the components of the GS energy $E$ are
\begin{eqnarray}
T[\gamma_1]=\int d1 \ t(\mathbf{r}_1)\gamma_1(1|1')\big|_{1'=1}, \quad
t(\mathbf{r}_1)=-\frac{1}{2}\frac{\partial^2}{\partial \mathbf{r}_1^2},
\nonumber \\
V_{\text{ext}}[\rho_1]=\int d1 \ \rho_1(1)v_{\text{ext}}(\mathbf{r}_1), \quad
V_{\text{int}}[\rho_2]=\int \frac{d1d2}{2!}\ \rho_2(1,2)v_{\text{int}}(r_{12}).
\end{eqnarray}
These are linear functionals. The generalized Hartree and Fock components of the
interaction energy $V_{\text{int}}$ are the bilinear functionals
\begin{equation}
V_{\text{H}}[\rho_1]=\int \frac{d1d2}{2!}\ \rho_1(1)\rho_1(2)
v_{\text{int}}(r_{12}), \;
V_{\text{F}}[\gamma_1]=\int \frac{d1d2}{2!}\ \gamma_1(1|2)\gamma_1(2|1)
v_{\text{int}}(r_{12}).
\end{equation}
Furthermore, the cumulant partitioning needs the linear functionals
\begin{eqnarray}
V_{\text{FC}}[w_2]= -\int \frac{d1d2}{2!}\ w_2(1,2) v_{\text{int}}(r_{12}),\;
V_{\text{C}}[u_2]= -\int \frac{d1d2}{2!}\ u_2(1,2) v_{\text{int}}(r_{12}).
\end{eqnarray}
Then the interaction energy is given by
\begin{equation}
V_{\text{int}}[\rho_2]=V_{\text{H}}[\rho_1]+V_{\text{FC}}[w_2], \quad
\rho_1=\frac{\int d2\rho_2}{N-1}, \quad w_2=\rho_1^2-\rho_2
\end{equation}
or
\begin{equation}
V_{\text{int}}[\gamma_1,u_2]=V_{\text{H}}[\rho_1]+V_{\text{F}}[\gamma_1]+
V_{\text{C}}[u_2], \quad \rho_1=\text{D}\gamma_1=\frac{\int d2u_2}{c_2}.
\end{equation}
Consequently, the total GS energy is
\begin{equation}
E[\gamma_2]=T[\gamma_1]+V_{\text{ext}}[\rho_1]+V_{\text{H}}[\rho_1]+
V_{\text{FC}}[w_2]
\end{equation}
or
\begin{equation}
{\tilde E}[\chi_2]=T[\gamma_1]+V_{\text{ext}}[\rho_1]+V_{\text{H}}[\rho_1]+
V_{\text{F}}[\gamma_1]+V_{\text{C}}[u_2].
\label{Etilde}
\end{equation}
The $N^2$ terms of $V_{\text{ext}}$ and $V_{\text{H}}$ cancel each other. In the
sum $V_{\text{H}}+V_{\text{F}}$, the self-interaction terms cancel each other.
In the sum
$V_{\text{F}}+V_{\text{C}}$ there is also a certain cancellation as it may be
seen from the normalization $\text{Tr}(\gamma_1^2+u_2)=N$, which does not
contain the L\"owdin parameter $c_2$. To find the GS, one has to minimize these
functionals $E[\gamma_2]$ or ${\tilde E}[\chi_2]$, provided that the
$N$-representability \cite{Cole} is taken into account: $\gamma_2$ follows from
$\gamma_3$, $\gamma_3$ from $\gamma_4$ etc. by successive contractions, cf.
Eq.(\ref{gammas}).

Other functionals implicitly defined by means of Levy's constrained search
\cite{Levy2} are (i) for the well tried and widely used DFT
\begin{equation}
\left(T+V_{\text{int}}\right)[\rho_1]=\min_{\Psi\rightarrow \rho_1}
\langle \Psi|\hat{T}+\hat{V}_{\text{int}}| \Psi \rangle, \quad
\rho_1=\text{D}\gamma_1, \quad
\gamma_1=\frac{\text{C}\gamma_2}{N-1}, \cdots ,
\end{equation}
from which follows $\left(T+V_{\text{FC}}\right)[\rho_1]=
\left(T+V_{\text{int}}\right)[\rho_1]-V_{\text{H}}[\rho_1]$, (ii) for the
DMFT (under discussion)
\begin{equation}
{\tilde V}_{\text{int}}[\gamma_1]=\min_{\Psi\rightarrow \gamma_1}
\langle \Psi|\hat{V}_{\text{int}}| \Psi \rangle, \quad
\gamma_1=\frac{\text{C}\gamma_2}{N-1}, \cdots ,
\end{equation}
from which follows ${\tilde V}_{\text{C}}[\gamma_1]=
{\tilde V}_{\text{int}}[\gamma_1]-V_{\text{H}}[\rho_1]-V_{\text{F}}[\gamma_1]$,
and (iii) for a possible pair-density functional theory (PDFT)
\begin{equation}
{\tilde T}[\rho_2]=\min_{\Psi\rightarrow \rho_2}\langle \Psi|\hat{T}| \Psi \rangle, \quad
\rho_2=\text{D}\gamma_2, \quad \gamma_2=\frac{\text{C}\gamma_3}{N-2}, \cdots.
\end{equation}
The 1-matrix functional ${\tilde V}_{\text{int}}[\gamma_1]$ has the simple scaling property
${\tilde V}_{\text{int}}[\gamma_1^{\lambda}]=\lambda {\tilde V}_{\text{int}}[\gamma_1]$, where
$\gamma_1^{\lambda}(\mathbf{r}_1|\mathbf{r}_1')=\lambda^3\gamma_1^{}(\lambda\mathbf{r}_1|\lambda\mathbf{r}_1')$.
The PD functional ${\tilde T}[\rho_2]$ has the simple scaling property
${\tilde T}[\rho_2^{\lambda}]=\lambda^2 {\tilde T}[\rho_2]$, where
$\rho^{\lambda}_2(\mathbf{r}_1,\mathbf{r}_2)=\lambda^6\rho_2^{}(\lambda\mathbf{r}_1,\lambda\mathbf{r}_2)$.
The reason of these simple scaling properties are the homogeneities of the corresponding
operators $\hat{V}_{\text{int}}$ and $\hat{T}$ with degrees $-1$ and $-2$, respectively. These
different degrees cause the more complicated scaling of the density functional
$(T+V_{\text{int}})[\rho_1]$, where the constrained search involves both
$\hat{T}$ and $\hat{V}_{\text{int}}$ \cite{Levy3}. To find approximate
functionals for
${\tilde V}_{\text{int}}[\gamma_1]$ and for ${\tilde T}[\rho_2]$, one may search for approximate functionals
$\chi_2[\gamma_1]$ (as e.g. proposed by K. Yasuda \cite{Rus}) and
$\gamma_1[\rho_2]$, respectively. From the above
introduced functionals follow the density functional
$E_{\rm DFT}[\rho_1]=(T+V_{\rm int})[\rho_1]+V_{\rm ext}[\rho_1]$, the
density-matrix (1-matrix) functional
$E_{\rm DMFT}[\gamma_1]=T[\gamma_1]+V_{\rm ext}[\rho_1]+
{\tilde V}_{\rm int}[\gamma_1]$, and the PD functional
$E_{\rm PDFT}[\rho_2]={\tilde T}[\rho_2]+V_{\rm ext}[\rho_1]+
V_{\rm int}[\rho_2]$, to be minimized, cf. Sec.VI(c),(d),(e).

\section{Basic equations}
\label{section4}

In order to find the GS $\Psi(1,\cdots,N)$, the starting point is the
MB Schr\"odinger equation  $\hat{H}\Psi=E\Psi$.
>From it follows the hierarchy of CSEs \cite{Cho} by multiplying it with
$\Psi'^{\ast}=\Psi^{\ast}(1',\cdots,N')$ and performing successively the $N-1$
contraction, the $N-2$ contraction etc.:
\begin{eqnarray}
\int d2 \cdots \left(\hat{H}-E\right) \Psi\Psi'^{\ast}\big|_{2'=2,3'=3,\cdots}=
0,\quad
\int d3 \cdots \left(\hat{H}-E\right) \Psi\Psi'^{\ast}\big|_{3'=3,4'=4,\cdots}=
0, \cdots .
\label{CSE}
\end{eqnarray}
On the lhss the RDMs $\gamma_1, \gamma_2, \gamma_3, \cdots$ appear. Thereby the CSE for
$\gamma_1$ contains $\gamma_2$ and $\gamma_3$ (1st formula), therefore it is called 1,3-CSE;
the CSE for $\gamma_2$ contains $\gamma_3$ and $\gamma_4$ (2nd formula), called
2,4-CSE, $\cdots$.
Expressing these RDMs by their cumulants $\chi_2, \chi_3, \cdots$
an equivalent hierarchy appears (no longer containing $E$); for a uniform notation one
should define the 1-body `cumulant' as $\chi_1\equiv\gamma_1$.

An equivalent writing of the Schr\"odinger equation is
$E=\min\limits_{\Psi}\langle\Psi|\hat{H}|\Psi\rangle, \; \langle\Psi|\Psi\rangle=N!$.
An equivalent writing of the hierarchy (\ref{CSE}) is
\begin{equation}
E=\min_{\gamma_2}E[\gamma_2],\quad
\gamma_2=\frac{\text{C}\gamma_3}{N-2}, \cdots,
\quad \text{Tr}\gamma_i=N(N-1)\cdots (N+1-i).
\label{equvi2}
\end{equation}
The size-extensive version of Eq.(\ref{equvi2}) is
$e=\min\limits_{\chi_2}{\tilde e}[\chi_2], \; \chi_2={\tilde {\rm C}}\chi_3,
\cdots, \; \text{tr}\gamma_i=c_i$ with ${\tilde e}[\chi_2]={\tilde E[\chi_2]}/N$, cf.
Eq.(\ref{Etilde}). It survives the TDL ($N,\Omega\to\infty, N/\Omega={\rm const}$).

Using the relation $\gamma_1={\tilde {\text{C}}}\chi_2$ described after
Eq.(\ref{rho2}), the total energy becomes a functional of $\chi_2$, which in its
spectral resolution provides geminals $\psi_K$ and occupancies $\mu_K$ as
variational parameters. Does this procedure give a link to the
Kimball-Overhauser approach?

\section{Size-extensitivity and thermodynamic limit}
\label{section5}

Solid state theory considers extended systems, which result from finite systems
($N$ electrons confined in a volume $\Omega$) through the TDL. In the terms of
thermodynamics one
has to distinguish intensive quantities, e.g. the density $\rho_1(1)$, and
extensive quantities, which are proportional to $N,$ like the normalization of
$\rho_1$ or the kinetic energy $T$. Amongst the quantities listed above are
also such ones, which contain terms proportional to $N^2$. Examples are
$V_{\text{ext}}$ and $V_{\text{H}}$. Fortunately their sum is size-extensive,
what makes the total energy $E$ also size-extensive. Thus the ratio $e=E/N$
survives the TDL, it defines the bulk
energy. Another example is the normalization of the PD $\rho_2$, where also
$N^2$ terms are present. This deficiency is `removed'
by considering the cumulants of $\rho_2$. Both $w_2(1,2)$ and $u_2(1,2)$ are
size-extensively normalized and the cumulant 2-matrix $\chi_2$ is
size-extensively contractable.

What happens with the equation $\hat H\Psi=E\Psi$ in the TDL? One may divide it
by $N$ and use the bulk energy $e=E/N$ on the rhs, but $\hat{H}/N$ has no
meaning in the TDL. Fortunately the hierarchy (\ref{CSE}) [being finite for
finite $N, \Omega$] only becomes infinitely long in the TDL. This
hierarchy has to be solved at least approximately. Even if this has been done
yielding the GS properties $\chi_2$ (and from it also $\gamma_1, \rho_1,
\rho_2$) and $e$, the question arises, how to obtain such important solid state
properties as the quasi-particle
band structure with its band gap and particle-hole bound states
(in the case of a semiconductor) and its Fermi surface and plasmons (in the
case of metals) and its effective masses and life-times. This task goes beyond
RDM theory. It needs the Green's function based QPT, cf. Sec.VI(i).

The TDL let emerge solid state properties having no
counterparts in small clusters. Such properties are long-range ordering phenomena
like ferromagnetism or other strong-electron-correlation phenomena like
metal-insulator transitions, Wigner-charge ordering, coexistence of ferromagnetism
and superconductivity, heavy fermions, quantum criticality, cf. Appendix.
\section{Classification of calculational methods}
\label{section6}

Amongst the calculational methods one may distinguish methods for finite
systems (atoms, molecules, clusters) and for extended systems (crystalline
solids) including jellium models. Another classification is: methods based on
the many-body WF and non-WF based methods, cf. Table \ref{tab:2nd}.

\begin{table}[h!]
\caption{\label{tab:2nd}
Methods to calculate the electronic structure.
The symbols (!), (?), (??) mean `well tried', `option', `idea', respectively.}
\begin{center}
\begin{tabular}{|c|c|c|c|}
\cline{2-3}
   \multicolumn{1}{c|}{}   &  finite        & extended           &
\multicolumn{1}{|c}{}  \\
\cline{2-3}
\hline
WF           & CI,CC,MP(!)     & $\rightarrow$ ?    & (a)\\
based        & QMC(!)            & $\rightarrow$ ?    & (b)\\
\hline
             & DFT(!)            & DFT(!)                & (c)\\
             & DMFT(?)           & DMFT(?)               & (d)\\
non-WF       & PDFT(??)           & PDFT(??)               & (e)\\
\cline{2-4}
based        & Coleman(?)     & $\to$ ?              & (f)\\
             & CSE(?)            & $\rightarrow$ ?    & (g)\\
             & ? $\leftarrow$ & Kimball-Overhauser & (h)\\
\cline{2-4}
             & QPT(!)            & QPT(!)                & (i)\\
\hline
\end{tabular}
\end{center}
\end{table}

\begin{enumerate}
\item[(a)] There are recent attempts to make the well-tried quantum-chemical
methods, e.g. CC, applicable also to crystalline solids. This subject was
thoroughly discussed at \cite{con3}, especially by P. Fulde, B. Paulus, and
U. Birkenheuer (incremental method). With this method (based on localized
Hartree-Fock  orbitals and a `constrained'
CI scheme), GS properties of insulating solids were calculated \cite{Ful3}. One
extension of the incremental scheme concerns metals and strongly correlated
systems. Thereby the problems appear of how to construct local orbitals in
metals, how the incremental scheme works with these orbitals and how to
handle the high degeneracy near the Fermi edges (with a MR method) \cite{PauRo}.
This allows
the method to be applied to the strongly correlated dissociation limit and
describe a metal-insulator transition \cite{Pau}. Another extension of the
incremental scheme aims to compute correlated wave functions and energies of
both valence and conduction bands \cite{Grae,Alb}. Thus the band
gap for periodic insulating systems including the effect of electron correlation
can be determined. First results are available for diamond and polyacetylene
\cite{Birk}.

R.J. Bartlett says ``CC theory offers the natural vehicle for transferring
1st-principle electronic structure informations into material modeling", in
[3b], p. 219. Because electron correlation is `short-sighted' (W. Kohn
\cite{Kohn}) a localized correlation treatment is developed using natural bond
orbitals \cite{Bart3}.

\item[(b)] For the QMC method in general cf. \cite{Ohno}. In order to reliably
predict materials properties, a QMC method using Hubbard-Stratonovich auxiliary
fields has been developed \cite{Zhang}. This MB method is aimed at treating
electron correlations in real materials. A first application concerns bulk Si.
Slater determinants from DFT calculations were used as the trial WFs.
The cohesive energy is comparable to (or better than) the best existing
theoretical results.
\end{enumerate}

\noindent
In contrast to the WF based methods (a) and (b), the methods (c)-(i) are
motivated by asking ``Is the WF not a monster containing too much
information, why one should struggle with it? Let us better concentrate on the
lowest order RDMs, namely $\gamma_1, \gamma_2$ and their positivity properties
and more general their $N$-representability (Coulson's challenge)", \cite{Cole}.

\begin{enumerate}
\item[(c)] DFT \cite{Hoh} uses the density $\rho_1$ [which has its charme and
power \cite{Levy}] as basic quantity. In its Kohn-Sham (KS) version,
the non-idempotent 1-matrix $\gamma_1$ is the sum of an idempotent
part $\gamma_1^{\text{KS}}$, corresponding to a non-interacting system with the
kinetic energy $T_{\text{KS}}[\rho_1]$, and a remainder, such that (i)
$\text{D}\gamma_1^{\text{KS}}= \sum_{\kappa}^{\rm occ}|\psi_{\kappa}(1)|^2$
(these KS orbitals $\psi_{\kappa}$ are not natural orbitals!) gives the correct
GS density $\rho_1$ of the interacting system (although $\gamma_1^{\rm KS}$ is
not the correct 1-matrix), (ii) the equation $(T+V_{\text{int}})[\rho_1]=
T_{\text{KS}}[\rho_1]+E_{\text{XC}}[\rho_1]$ defines the exchange-correlation
(XC) functional $E_{\text{XC}}[\rho_1]$ to be approximated (LDA, GGA, $\cdots$),
and (iii) the minimization of $E_{\rm DFT}[\rho_1]$ yields an effective 1-body
Schr\"odinger equation with an XC potential $v_{\rm XC}(1)=
\delta E_{\rm XC}[\rho_1]/\delta \rho_1(1)$. Finally, its solution gives
(together with the aufbau principle) the density $\rho_1(1)$
and the total energy $E$.
Besides Lagrange multipliers $\varepsilon_{\kappa}$ appear, which
surprisingly approximate the quasi-particle band structure of sp metals as
(from a conceptional point of view) illegitimate children of the KS-DFT (W. Kohn).
But for semiconductors their KS-DFT band gaps are too small and for heavy-fermion
systems the experimental effective masses are much larger than the KS-DFT
ones. DFT goes beyond the correlation neglecting Hartree-Fock approximation.
It is well-tried and widely used for both finite and extended (metallic and
semiconducting/insulating) systems,
cf. e.g. \cite{Parr}. The interconnections between DFT and CC theory are
studied (R.J. Bartlett at \cite{con5}). Activities are going on trying to
remove deficiencies and failures of this method (e.g. the van-der-Waals
interaction \cite{Ko3}, the problems listed under (d), the asymptotic behavior
of the XC potential \cite{Niq}), and to generalize it, e.g. DFT for MR cases
(A. Savin at \cite{con5}), time-dependent DFT \cite{Bur} and its relation to
QPT \cite{Oni}, DFT for excited states \cite{Nagy}, DFT for transport properties
\cite{Zac}. In particular the originally used LDA [based on the HEG
and being the Thomas-Fermi approximation on a higher level] is
modified (i) by climbing the `Jacob's ladder' via GGA and Meta-GGA up
to Hyper-GGA, which are non-empirical approximations for
$E_{\text{XC}}[\rho_1]$ designed for molecules and solids
\cite{Tao}, (ii) by studying orbital functionals (optimized effective potential
method, the DF method in 3rd generation), cf. e.g. \cite{Kuem}, and
(iii) by taking into account peculiarities of strong correlation (Appendix),
which suppresses charge fluctuations, thus localizes electrons and therefore
favors insulating behavior. Strong electron correlation is often treated with
the help of the Hubbard model (containing the on-site Coulomb repulsion $U$),
the $t$-$J$ model, the Anderson model (including attempts to determine model
parameters from 1st principles \cite{Czy}) or the orbital model \cite{Tjeng} and
the density-matrix renormalization group \cite{Noa}. One way for (iii) is the
combination
with the Hubbard model (LDA+U); its possible link to 1-body Green's function
theory is discussed in Ref.\cite{Esch}, where a density functional application
to strongly correlated electron systems is presented. Another way is the
combination with the dynamical mean field theory of strongly correlated
fermion systems \cite{Geor}. This allows realistic calculations of transition
metal oxides and f-electron materials, e.g. the Mott-Hubbard transition of
$\text{V}_2\text{O}_3$, the volume collapse transition of Ce,
the magnetic properties of LiV$_2$O$_4$, and the photoemission spectra of
SrVO$_3$ and CaVO$_3$ \cite{Held}.
\item[(d)] DMFT uses the 1-matrix $\gamma_1$ as basic quantity \cite{Gil}.
Unlike DFT, it does not need a (non-interacting) reference system, but it needs
an approximation for ${\tilde V}_{\text{int}}[\gamma_1]$ or
${\tilde V}_{\text{C}}[\gamma_1]$, from which follows
the functional $E_{\rm DMFT}[\gamma_1]$. Its minimization
yields $\gamma_1$ and $E$ (but not $\rho_2$ and in the case of solids
not the band structure), cf. e.g. \cite{Goed}. Each of the so far proposed
approximations have their successes and shortcomings (e.g. violation of the
positivity condition or of the contraction SR). Conditions for
approximate 1-matrix functionals are in Ref.\cite{Cio2}. From a conceptional
point of view, DMFT is simpler than DFT. The hope is that the new method
is like DFT cheaper than CI, CC, MP and applies to both
finite and extended (metallic and insulating) systems and that it can handle
problems where DFT fails \cite{Goe2}:
\begin{itemize}
\item negative atomic ions are not bound, ionization potentials are too small,
\item correlation energies of He($Z$) and Be($Z$) have not the correct large-$Z$
behavior, cf.\cite{Dav2},
\item errors in the bond lengths and bond angles of some
molecules, e.g.
$\text{CH}_4, \text{H}_2\text{O}, \text{BH}, \text{O}_3, \text{Li}_2,
\text{Be}_2$ (through this
series, static correlation is progressively increasing),
\item the reaction $\text{H}_2+\text{H}\rightarrow \text{H}+\text{H}_2$  is
another MR case, it has no transition state in LDA and one that is too low with
better functionals beyond LDA,
\item wrong dissociation limits,
\item cohesive energies of bulk metals are overestimated in LDA and
underestimated in GGA,
\item DFT poorly performs for polyacetylene fragments and oligoporphyrins,
incorrectly predicting a triplet GS for polyacetylene \cite{Cai}.
\end{itemize}
There are links between DMFT and the RDM theory (reconstruction of the 2-matrix)
\cite{Herb1} and between DMFT and the theory of antisymmetrized products of
strongly orthogonal geminals \cite{Cio3}. For the geminal theory in general cf.
\cite{Sur}.
\item[(e)] The possible PDFT uses the PD $\rho_2$ as basic quantity. Again
unlike DFT, it does not need a (non-interacting) reference system, but it needs
an approximation for ${\tilde T}[\rho_2]$, from which follows the functional
$E_{\rm PDFT}[\rho_2]$. Its minimization together with an ansatz for $\rho_2$ in
terms of geminals should yield $\rho_2$ and $E$ for both
finite and extended systems (but not $\gamma_1$ and in the case of solids not
the band structure) \cite{Zie2}. - Another
method, which uses also two-particle states (what involves the extension of LDA
from densities to PDs), is in Ref.[3b], p. 325 and refs. therein.
\item[(f)] For the RDM theory in general cf. \cite{Dav1,Erd,Cio1,Cole}. For the
investigation of the $N$-representability in terms of Kummer variety,
Coleman's algorithm, and linear inequalities cf. Ref.\cite{Cole}. Whether this
way is really cheaper than a full CI calculation and whether it applies to
extended systems has to be checked.
  \item[(g)]The generation of higher-order RDMs from $\gamma_2$ via
`reconstruction'
has been developed by Nakatsuji, Valdemoro, and Mazziotti \cite{Cio1}, p. 85,
117, and 139, respectively, and refs. therein, cf. also \cite{Nakuji,Nak,
Mazz}. The philosophy behind this approach is: Let us use an exact
functional and violate slightly $N$-representability rather than using
approximate DFT functionals. One may criticize this attitude with the
argument: Already a slight violation of the $N$-representability opens regions
of the variational space, where the calculated energy
is below the GS energy \cite{Harr}. There seem to be
encouraging results: using the positivity condition on the 2-matrix
(as a necessary condition of the $N$-representability), energies of closed- and
open-shell atoms and molecules were calculated variationally \cite{Nak} and
energies of atoms and molecules (both near and far from equilibrium geometries)
were calculated solving the 1,3-CSE \cite{Mazz}. But critical discussions
concern `how the results change, when going from minimal to extended basis
sets?' and `are the 3- and/or 4-cumulants really negligible?' \cite{Harr, Herb,
Noo}. Whether the CSE method (provided that it survives these criticisms) applies also to
extended systems has to be checked.
\item[(h)] In the Kimball-Overhauser approach \cite{Kim} the singlet PD $g_+(r)$
and triplet PD $g_-(r)$ of the HEG (here normalized as $g_\pm(\infty)=1$) are
parametrized in terms of geminals $R_l(r,k)$, which are intuitively solutions of
\begin{equation}
\left[
-\frac{1}{r}\frac{\partial^2}{\partial r^2}r+\frac{l(l+1)}{r^2}+\frac{1}{r}+v_{\text{scr}}^{\pm}(r)-k^2
\right]R_l(r,k)=0,
\label{2bodySch}
\end{equation}
such that $g(r)=\frac{1}{4}\left[ g_+(r)+3g_-(r)\right]$
with
\begin{equation}
g_{\pm}(r)=2\sum_{L}^{\pm}\frac{2}{N}\sum_{\mathbf{k}}\mu(k)R_l^2(r,k), \quad
\mu(k)=\frac{2}{N}\sum_{\mathbf{K}}n(k_1)n(k_2),
\label{PDs}
\end{equation}
where $k_1=\big|\frac{1}{2}\mathbf{K}+\mathbf{k}\big|, k_2=\big|\frac{1}{2}
\mathbf{K}-\mathbf{k}\big|$ , $L=(l,m_l)$, $+$ is for even $l$ and $-$ for odd
$l$. $v_{\text{scr}}^{\pm}(r)$ describes the effective screening of the Coulomb
repulsion by the Fermi-Coulomb hole around each electron; on the Hartree level
it follows from $\triangle v_{\text{scr}}^{\pm}(r)=4\pi\rho[1-g(r)]$. The $R_l$ are
scattering states, characterized by scattering phase shifts. In Ref.\cite{Zie5}
the normalization and contraction SRs of the 2-matrix is related to these phase
shifts. According to these new SRs it should be possible to calculate not only
the PD from $v_{\rm scr}(r)$, but also the momentum distribution $n(k)$.
For further details cf. \cite{Kim,Davou,Zie5}
and refs. therein. - Can Eq.(\ref{2bodySch}) be derived from MB theory
(CSE, contraction SRs, Bethe-Salpeter equation or other methods)? - To what extent is the
Kimball-Overhauser approach related to the PDFT (e) and to the 2-body cluster
expansion of
Ref.\cite{Pie}? - Whether this method correspondingly modified applies also to
inhomogeneous and also to finite systems has to be studied.
- A general remark of E.R. Davidson is ``Scientists
have not yet learned to think in terms of $\binom{N}{2}$ geminals rather than in
$N$ orbitals'' \cite{Dav1}, p. 97.
\item[(i)] The Green's function based QPT (like the KS-DFT) rests upon an
effective 1-body
Schr\"odinger equation with the local, real and energy independent XC
potential $v_{\text{XC}}(1)$ of the KS equation replaced by the non-local,
complex and energy dependent XC part $\Sigma_{\text{XC}}(1|1',\varepsilon)$
of the self-energy \cite{Hed}. There holds an integral relation between
$v_{\rm XC}$ and $\Sigma_{\rm XC}$ \cite{Sham2}, Eq.(11). QPT needs an
approximation for the
$\Sigma_{\text{XC}}$ (e.g. the GW approximation and its vertex correction) and
yields the GS properties $\gamma_1$ and $E$ and besides the quasi-particle band
structure $\varepsilon_k$. Total energies were calculated (\cite{Ani}, p. 85)
with the Galitskii-Migdal formula \cite{Holm} and with the Luttinger-Ward
formula \cite{Miy}. The real part of $\Sigma_{\text{XC}}$ compensates
(in the case of semiconductors) for the otherwise (namely within DFT)
underestimated band gaps and yields mass renormalizations.
The imaginary part of $\Sigma_{\text{XC}}$ accounts for life-time effects.
The Coulomb interaction between electrons in the conduction band and holes in
the valence band are included by solving the Bethe-Salpeter equation for the
electron-hole two-particle Green's function \cite{Sham}. This provides the
excitonic binding energies \cite{Pusch}. The combination with the afore already
mentioned dynamical mean field theory allows the first-principle approach to
strongly correlated systems (Sec. VI(c), Appendix) \cite{Bier,Kats}.
\end{enumerate}
\section{Summary}
\label{summary}

Summarizing and focusing on solids, there are several methods to calculate such
GS properties as $E, \rho_1, \gamma_1$, but from a conceptional point of view
only QPT Sec.VI(i) and the incremental method VI(a) yield also the band structure
including Fermi surfaces (of metals) and band gaps (of semiconductors)
and effective masses. So far a weak point is the PD $\rho_2$, cf. Sec.VI(e)-(h).
This weakness can be removed, if the cumulant 2-matrix $\chi_2$ is calculated.
This is a key quantity because from it follows: the 1-matrix $\gamma_1$ (which
contains the density $\rho_1$) and using this result, the 2-matrix (which
contains the PD). This PD is needed to calculate the interaction
energy and particle number fluctuations in parts of the system (the magnitude
of which measures the correlation strength). The spectral resolution of
$\chi_2$ leads to geminals, which may be the
solutions of a 2-body Schr\"odinger equation with a screened Coulomb interaction
if not exact, but possibly to a good approximation. What peculiarities has
this screening potential for different types of solids like sp metals,
transition metals, semiconductors/insulators, superconductors, 
ferro/antiferro-magnets, strongly correlated systems? 
Further work is called for to answer the questions posed in this paper.

\begin{acknowledgments}
The authors thank A. Beste, U. Birkenheuer, P. Fulde, S. Goedecker, J.M.
Herbert, W. Koppler, M. Laad, B. Paulus, J. Richter, M. Sigrist, C. Umrigar, M. Vojta for helpful
discussions and
A. Tchougreeff for organizing the V.A. Fock School on Quantum and Computational
Chemistry (Velikiy Novgorod, 12-16 May 2003). P.Z. and F.T.
thank P. Fulde and H. Eschrig, respectively, for supporting this work.
\end{acknowledgments}

\appendix
\section*{Appendix: Correlation strength and strong correlation}
\label{section7}

The strength of correlation is not a matter of the correlation energy. More
relevant is the quantum-kinematics expressed in terms RDs and RDMs. So the
L\"owdin parameter $c_2$ is one correlation index, cf. Eq.(\ref{Low}).
Electron-number fluctuations $\Delta N_{X}$ in a part $X$ of the system are
another index, decreasing with increasing correlation, i.e.
correlation suppresses such particle-number fluctuations \cite{Ful1,Zie8}, it
thus favors electron localization and non-metallic behavior.
The Calogero-Sutherland model shows that $\Delta N_{X}$ is more sensitive,
because it distinguishes between attractive and repulsive interaction, whereas
$c_2$ shows practically no change when
changing the sign of the interaction (the non-idempotency of the momentum
distribution is qualitatively the same for these two cases) \cite{Roe}.

Cases of strong correlation have to be described by MR states,
cf. the discussion after Eq.(\ref{Low}). This strong-correlation case is
associated with a strong suppression of fluctuations or equivalently with
strong localization, cf. discussion after Eq.(\ref{PX1}). Examples are:
(i) He($Z$) for $Z{_> \atop ^{\to}} Z_{\text{cr}}=0.91$,
where one electron is localized at the nucleus while the
other one becomes unbound and escapes \cite{Zie9},
(ii) the $\text{H}_2(R)$ molecule in the dissociation limit
$R\rightarrow\infty$, where at each nucleus one electron is localized such that
the total spin state is a singlet \cite{Zie10},
(iii) the Wigner crystallization of the low-density electron gas.
For the case of `strictly correlated' systems cf.\cite{Seidl}. Ferro- and
antiferro-magnetism and corresponding quantum phase transitions are `old'
strong-correlation phenomena. `More new' examples are:
\begin{itemize}
\item Mott insulators, where the hopping
between the lattice sites is suppressed by a large Hubbard $U$:
CuO, NiO, CoO \cite{Fri}, FeO, MnO, CaCuO$_2$, V$_2$O$_3$, Ca$_2$RuO$_4$, 
La$_2$CuO$_4$. Examples for correlation-driven 
metal-insulator transitions are FeSi (under doping), LaCoO$_3$ (under temperature), 
V$_2$O$_3$ and Ca$_2$RuO$_4$ (under pressure).
\item Charge ordering in Yb$_4$As$_3$ or NaV$_2$O$_5$,
charge-density-wave transition in CuV$_2$S$_4$, spin-density waves in
Ce(Ru$_{1-x}$Rh$_x$)$_2$Si$_2$, charge-spin separation in
SiCuO$_2$, Yb$_4$As$_3$ (the thereby appearing
low-lying excitations are not quasiparticles, but collective spin and charge
density fluctuations).
\item Quasi-one-dimensional systems: charge-spin separation in
polyacetylen (solitons) or in Bechgaard salts (heat transport) \cite{Lor},
spin-Peierls transition in CuGeO$_3$, non-Fermi liquid behavior in organic
conductors.
\item High $T_{\rm c}$ superconductors and other unconventional superconductors
[Sr$_2$RuO$_4$, ScCu(BO$_3$)$_2$]. $\text{UGe}_2$, \text{URhGe} are
superconducting ferromagnets \cite{Hux}.
\item Below characteristic temperatures heavy-fermion systems (mostly compounds 
of light lanthanoides and actinides with their 4f and 5f electrons) show anomalously 
large electronic heat capacities, which can be described as heavily renormalized 
quasiparticles. They have huge effective masses being several hundred times the 
free-electron mass. Examples are (P. Fulde, in [3b], p. 111):
$\text{CeAl}_3$, $\text{CeCu}_2\text{Si}_2$, and CeCoIn$_5$
(Kondo effect), $\text{Nd}_{2-x}\text{Ce}_x\text{CuO}_4$ (Zeeman effect),
CeCu$_{6-x}$Au$_x$; 4f holes (respectively spin
excitations) are decisive for $\text{Yb}_4\text{As}_3$  and 3d electrons for 
$\text{YMn}_2$, LiV$_2$O$_4$. The compounds
UPd$_2$Al$_3$ and $\text{UPt}_3$ are characterized by a combined coherence of
localized and itinerant electrons [the 5f electrons are partially localized as
in the sandwich molecule $\text{U}(\text{C}_8\text{H}_8)_2$], they are described
by a dual model of localized and hybridized 5f electrons.
The compounds CeCu$_2$Si$_2$, UBe$_{13}$, UPd$_2$Al$_3$, UPt$_3$, CeIn$_3$ are
superconductors, ZrZn$_2$ is a superconducting ferromagnet \cite{Hay}.
The compounds CeMIn$_5$ (M=Co, Rh, Ir) are antiferromagnets. The compounds
YbRh$_2$Si$_2$ and YbRh$_2$(Si$_{0.95}$Ge$_{0.05}$)$_2$ have small
antiferromagnetic ordering temperatures at 70 mK and 20 mK, respectively,
and show other peculiar low-temperature behavior \cite{Cust}.
\item Quantum criticality near $T=0$ (where quantum fluctuations predominate
thermal fluctuations) in CeCu$_{6-x}$Au$_x$, YMn$_2$, CePd$_2$Si$_2$,
CoCu$_2$Si$_{2-x}$Ge$_x$.
\end{itemize}
Other aspects are: mixed valence compounds ($\text{SmS}$, Yb$_4$As$_3$),
geometrical frustration (LiV$_2$O$_4$), frustrated spin systems, non-Fermi
liquid behavior (MnSi, CeCu$_2$Si$_2$, UPd$_2$Al$_3$, YbRh$_2$Si$_2$,
SrRuO$_3$ is a non-Fermi liquid ferromagnet),
fractionally charged excitations in the fractional quantum Hall effect with its
composite fermions (electrons+flux lines), `fractional charges' (topological
excitations of special low-temperature phases, possibly in LiV$_2$O$_4$). For
reviews on strong electron
correlation cf. Ref.\cite{Ani}. Theoretical treatments (ab-initio or with
models) are mentioned in Sec.VI(c),(i).

For solids, the bandwidth $W$ is compared with the Hubbard $U$,
then addressing $W\gg U$ as weak and $W{_< \atop ^{\approx}}U$ as strong
correlation with $W\ll U$ as the atomic limit. Question: How are $c_2,
\Delta N_{X}$, static/dynamic correlation, $W \gtrless U$ mutually related?


\end{document}